\documentclass[aps,pra,twocolumn,letterpaper,superscriptaddress,footinbib,showpacs]{revtex4-1}
\usepackage{times,graphicx,amsmath,amssymb,array,sidecap}
\usepackage[hyperfigures,colorlinks=true,allcolors=blue]{hyperref}
\usepackage{apjfonts}
\usepackage[utf8]{inputenc}
\usepackage{color}

\newcommand{\abs}[1]{\left|#1\right|}
\newcommand{\bra}[1]{\left\langle \, #1 \,\right|}
\newcommand{\ket}[1]{\left|\, #1 \, \right\rangle}
\newcommand{\bket}[2]{\left\langle \, #1 \,|\, #2 \, \right\rangle}
\newcommand{\boket}[3]{\langle\, #1 \,|\, #2 \,|\, #3 \,\rangle}
\newcommand{\pb}{\Phi_0}

\newcommand{\cj}{C_\text{J}}
\newcommand{\bcj}{\mathsf{C}_\text{J}}

\newcommand{\cs}{C_\Sigma}
\newcommand{\bcs}{\mathsf{C}_\Sigma}
\newcommand{\cp}{C}
\newcommand{\bcp}{\mathsf{C}}

\newcommand{\ec}{E_\text{C}}
\newcommand{\ecj}{E_{\text{CJ}}}
\newcommand{\ecs}{E_{\text{C}\Sigma}}
\newcommand{\ej}{E_\text{J}}

\newcommand{\dej}{\delta\!E_\text{J}}
\newcommand{\bdcj}{\,\delta\! {\mathsf{C}_\text{J}}\,}
\newcommand{\dc}{\delta\!C}
\newcommand{\dcj}{\delta\! {C_\text{J}}}
\newcommand{\bdc}{\delta\!\mathsf{C}\,}
\newcommand{\del}{\delta\!E_L}

\newcommand{\px}{\varphi_\text{ext}}

\newcommand{\be}{\begin{equation}}
\newcommand{\ee}{\end{equation}}

\begin{document}
\title{Understanding degenerate ground states of a protected quantum circuit in the presence of disorder}
\author{Joshua Dempster}
\affiliation{Department of Physics \& Astronomy, Northwestern University, Evanston, IL 60208, USA}
\author{Bo Fu}
\affiliation{Department of Physics \& Astronomy, Northwestern University, Evanston, IL 60208, USA}
\author{David G.\ Ferguson}
\altaffiliation[Current address: ]{Advanced Concepts \& Technologies Division, 
Northrop Grumman Corporation, Linthicum, Maryland 21090, USA}
\affiliation{Department of Physics \& Astronomy, Northwestern University, Evanston, IL 60208, USA}
\author{D.\ I.\ Schuster}
\affiliation{Department of Physics and James Franck Institute, University of Chicago, Chicago, Illinois 60637, USA}
\author{Jens Koch}
\affiliation{Department of Physics \& Astronomy, Northwestern University, Evanston, IL 60208, USA}
\date{\today}

\begin{abstract}
 A recent theoretical proposal suggests that a simple circuit utilizing two superinductors may produce a qubit with ground state degeneracy [P.\ Brooks \emph{et al.}, Phys.\ Rev.\ A \textbf{87}, 052306 (2013)]. We perform a full circuit analysis along with exact diagonalization of the circuit Hamiltonian to elucidate the nature of the spectrum and low-lying wave functions of this $0-\pi$ device. We show that the  ground state degeneracy is robust to disorder in charge, flux and critical current as well as insensitive to modest variations in the circuit parameters. Our treatment is non-perturbative, provides access to excited states and matrix elements, and is immediately applicable also to intermediate parameter regimes of experimental interest.
\end{abstract}

\maketitle

\section{Introduction}

The idea of topological protection from decoherence \cite{Kitaev2003} has greatly influenced research aimed at the physical implementation of quantum computation. The central paradigm of topological protection is to store quantum information in an explicitly non-local fashion, rendering qubits insensitive to various sources of local noise. Potential realizations of such topological protection have been suggested for anyon quasiparticles in fractional quantum Hall systems \cite{Kitaev2003,Nayak2008,Abanin2012}, for $p+ip$ superconductors \cite{Read2000}, as well as for Majorana fermions in topological nanowires \cite{Hassler2010,Alicea2011,Jiang2011,Hassler2011,VanHeck2012,Pekker2013}. With coherence times of the order of milliseconds \cite{Paik2011a,Rigetti2012}, conventional unprotected superconducting circuits \cite{Nakamura1999,Bouchiat1998,Devoret2004a,Clarke2008}  are  already quite promising \cite{Schoelkopf2008,Devoret2013a}.
Here we will attempt to explore what are the minimal requirements for such intrinsic protection in superconducting circuits. We will show that the use of circuits with more than one or two effective degrees of freedom can lead to qualitatively different and more robust quantum states.

A promising avenue for realizing protection in superconducting circuits is to exploit frustration, such as Ising-type in larger junction arrays \cite{Gladchenko2008,Doucot2009,Doucot2012a}. 
From this viewpoint,  the recently proposed $0-\pi$ circuit by Brooks, Kitaev and Preskill (BKP) \cite{Brooks2013} is particularly intriguing: it features a much smaller four-node superconducting circuit with the potential of remarkable robustness with respect to local noise and the possibility of carrying out quantum gates in a protected fashion.

The BKP paper takes it for granted that a $0-\pi$ qubit with sufficient inductance and without disorder can be realized, and rather focuses on protected gates. 
A challenge of realizing the circuit is that it requires inductances larger than have been realized. Nevertheless,  there has been significant experimental advances towards building such ``superinductors" \cite{Manucharyan2009,Manucharyan2010,Bell2012,Masluk2012}.
Here, we investigate what concrete device parameters are needed for robust degeneracy and discuss how realistic accessing this parameter regime is based on current knowledge and state-of-the-art fabrication techniques. In doing so, we further elucidate the nature of wave functions and spectral properties of the $0-\pi$ circuit and discuss in detail the effects of device imperfections, in particular disorder in circuit element parameters, on the characteristics of the $0-\pi$ circuit.

Our presentation is structured as follows. We start with the full circuit analysis of the $0-\pi$ device in Section \ref{sec:circuitanalysis}, and thereby identify its three relevant degrees of freedom. If disorder in device parameters is absent, the $0-\pi$ device is described by merely two degrees of freedom, while the third one decouples. For this ideal case, we investigate the spectrum, wave functions and the degeneracy of low-lying states in Section \ref{sec:idealspectrum}. The role of disorder in circuit parameters is addressed in Section \ref{sec:disorder}. We show that the $0-\pi$ circuit is favorably insensitive to disorder in junction parameters but may be negatively affected by disorder in the values of the superinductances and the additional capacitances in the circuit. We present our conclusions and an outlook on possible future uses of the $0-\pi$ circuit in Section \ref{sec:conclusions}.

\begin{figure}
\centering
	\includegraphics[width=0.5\columnwidth]{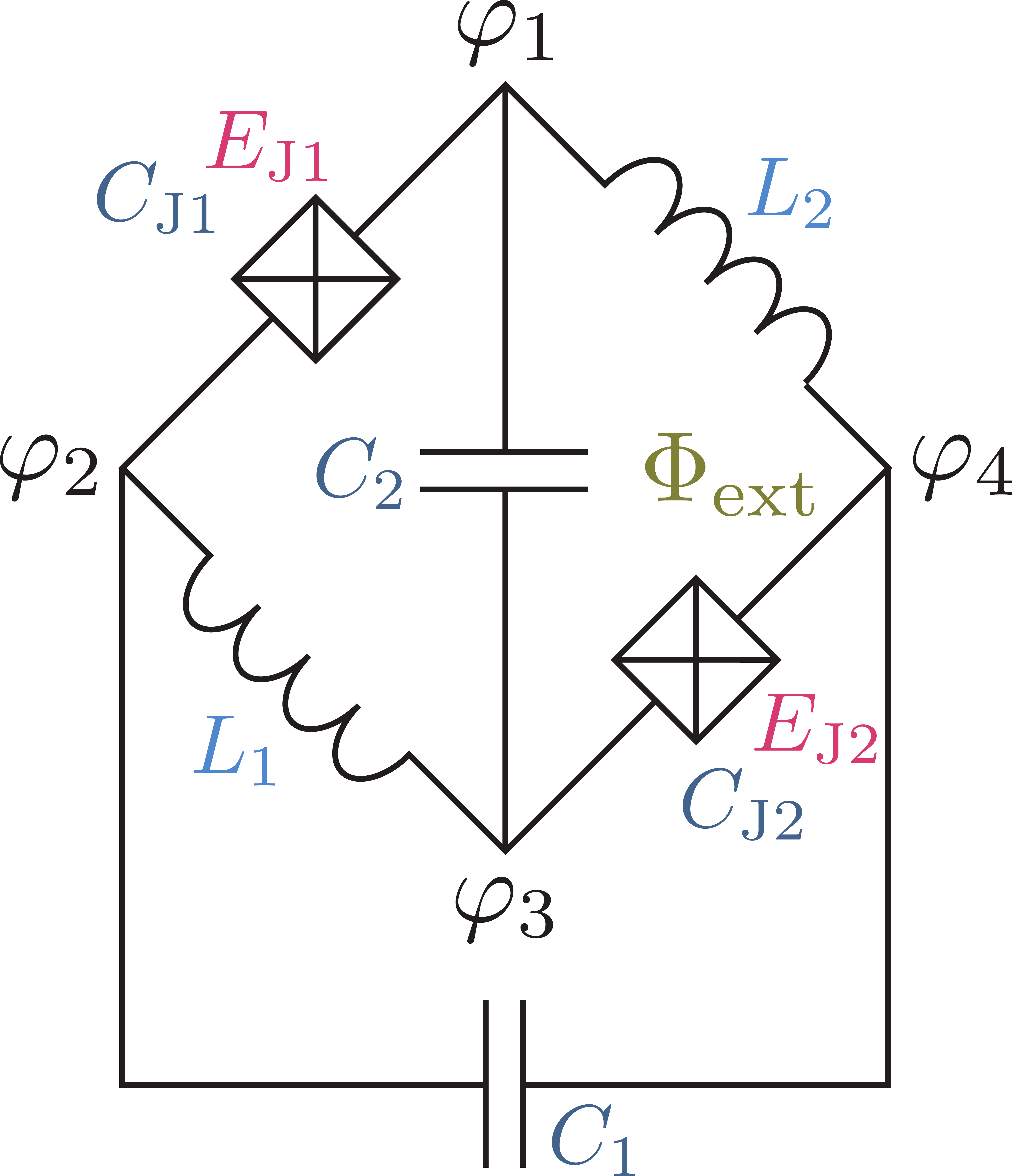}
	\caption{(Color online) Circuit diagram of the $0-\pi$ qubit. Each of the four circuit nodes is associated with one phase variable $\varphi_j$. Additional symbols show the naming of capacitances, inductances and Josephson junction parameters, and the magnetic flux $\Phi_\text{ext}$ that may thread the inner circuit loop.	\label{fig-circuit}}
\end{figure}

\section{Circuit Analysis of the symmetric $0-\pi$ device\label{sec:circuitanalysis}}
The $0-\pi$ device, depicted in Fig.\ \ref{fig-circuit}, is a superconducting circuit with four nodes. The nodes form an alternating ring consisting of two inductors and two Josephson junctions. Additional cross-capacitances connect the opposing nodes in the ring. As shown, all circuit elements occur pairwise  and are, in the ideal circuit, identical such that both Josephson junctions share the same Josephson energy $\ej$ and junction capacitance $\cj$, both inductances are given by $L$, and both cross-capacitances by $\cp$. Neglecting any deviations in these circuit element parameters renders the circuit symmetric under a $\pi$ rotation. (For better visibility, one cross-capacitance is shown external to the ring in Fig.\ \ref{fig-circuit}.) We hence refer to this special case as the \emph{symmetric} $0-\pi$ device. 

For the quantitative study of the spectrum and eigenstates of the symmetric $0-\pi$ device, we begin with a systematic circuit analysis.
In the usual first step  \cite{Devoret1995,Burkard2004}, we assign  node fluxes  
to each of the  four circuit nodes numbered $j=1,\ldots,4$. Each node flux, defined as the time integral of the electrostatic potential $U_j$ on each node,
 serves as a generalized variable in the circuit Lagrangian. For convenience, we employ the dimensionless version of the node variables, $\varphi_j =\int_{t_0}^t dt'\, U_j(t')
/\pb$, where $\pb=\hbar/2e$  is the reduced magnetic flux quantum. (Note the inclusion of the $1/2\pi$ factor relative to the conventional definition of the magnetic flux quantum).

Expressed in terms of node variables, the kinetic and potential energy contributions to the circuit Lagrangian assume the form
\begin{align}\label{Tkin}
T=&\tfrac{1}{2}\bcj(\dot\varphi_2-\dot\varphi_1)^2
+\tfrac{1}{2}\bcj(\dot\varphi_4-\dot\varphi_3)^2\\\nonumber
&
+\tfrac{1}{2}\bcp(\dot\varphi_3-\dot\varphi_1)^2
+\tfrac{1}{2}\bcp(\dot\varphi_4-\dot\varphi_2)^2
\end{align}
and
\begin{align}
U=&-\ej \cos(  \varphi_4-\varphi_3 -\px/2)
-\ej \cos(  \varphi_2-\varphi_1 - \px/2)\nonumber\\
&+\tfrac{1}{2}E_{L}(\varphi_2-\varphi_3)^2
+\tfrac{1}{2}E_{L}(\varphi_4-\varphi_1)^2.
\label{potenergy}
\end{align}
Note that we have absorbed factors of $\pb^2$ by letting $\bcp=C\pb^2$ etc.
The potential energy $U$ incorporates the effect of an external magnetic flux, expressed here in dimensionless form as $\px=\Phi_\text{ext}/\Phi_0$. We have chosen a symmetric division of the flux between the two Josephson junctions. (As usual, other equivalent choices are simply obtained by shifting the node variables.) The terms in the second line of Eq.\ \eqref{potenergy} denote the inductive contributions in terms of the energy $E_L=\pb^2/L$.

From Eq.\ \eqref{Tkin} it is clear that there will be cross-terms between the $\dot\varphi_j$ variables. Physically, such terms  arise because of  cross-capacitances in the circuit diagram, Fig.\ \ref{fig-circuit}. 
We now adopt new variables $\phi$, $\theta$, $\chi$, and $\Sigma$, which we will show to diagonalize the kinetic energy term and which are defined as
\begin{align}\label{transf1}
&2\phi=(\varphi_2-\varphi_3)+(\varphi_4-\varphi_1),\qquad
2\chi=(\varphi_2-\varphi_3)-(\varphi_4-\varphi_1),\nonumber\\
&2\theta=(\varphi_2-\varphi_1)-(\varphi_4-\varphi_3),\qquad
\Sigma= \varphi_1 +\varphi_2 + \varphi_3  + \varphi_4
\end{align}
with inverse
\begin{align}
2\varphi_1 &= \Sigma - \theta - \phi+\chi,
&2\varphi_2 = \Sigma + \theta + \phi+\chi,\\\nonumber
2\varphi_3 &= \Sigma + \theta - \phi-\chi,
&2\varphi_4 = \Sigma - \theta + \phi-\chi.
\end{align}

Following our variable transformation, the kinetic and potential energies simplify to
\begin{align}\label{kinetic}
T&=\bcj \dot\phi^2 + \bcs\dot\theta^2+\bcp\dot\chi^2
\end{align}
 and
\begin{align}
U=&-2\ej \cos\theta\,\cos(\phi-\px/2) +E_L\phi^2+E_L\chi^2.
\label{potential}
\end{align}
Here, $\bcs=\bcj+\bcp$ abbreviates the sum capacitance, again including the factor of $\pb^2$. As intended, the effective mass tensor in Eq.~\eqref{kinetic} is now diagonal.
Due to gauge invariance, the variable $\Sigma$  decouples completely, leaving us with  three  degrees of freedom. 
Furthermore, the variable $\chi$ is harmonic: it simply captures the oscillator subsystem with frequency $\Omega_\chi=\sqrt{8E_L \ec}/\hbar$ formed by the two inductances $L$ and the two capacitances $C$; $\ec=e^2/2C$ denotes the relevant charging energy.  For the perfectly symmetric circuit, the oscillator variable $\chi$  exactly decouples from the other two variables $\theta$ and $\phi$ but will become relevant again once we consider disorder in Section \ref{sec:disorder}.

For the remaining two degrees of freedom of the symmetric $0-\pi$ qubit we thus obtain the  effective Lagrangian
\be
\mathcal{L}_\text{sym}= \bcj\dot\phi^2 + \bcs\dot\theta^2+2\ej \cos\theta\cos(\phi-\px/2)-E_L\phi^2.
\ee
Note that, here,  $\dot\phi$ only sees the junction capacitance, whereas $\dot\theta$  sees (i.e., depends  on the phase difference across) both the junction as well as the other two cross-capacitances.  The new effective masses associated with them may thus be different and will be instrumental in understanding the physics of the circuit. From the potential energy terms it is clear that both $\phi$ and $\theta$ are affected by  the junctions,  but only $\phi$ is influenced by the inductors. 

\begin{figure*}
\centering
	\includegraphics[width=0.75\textwidth]{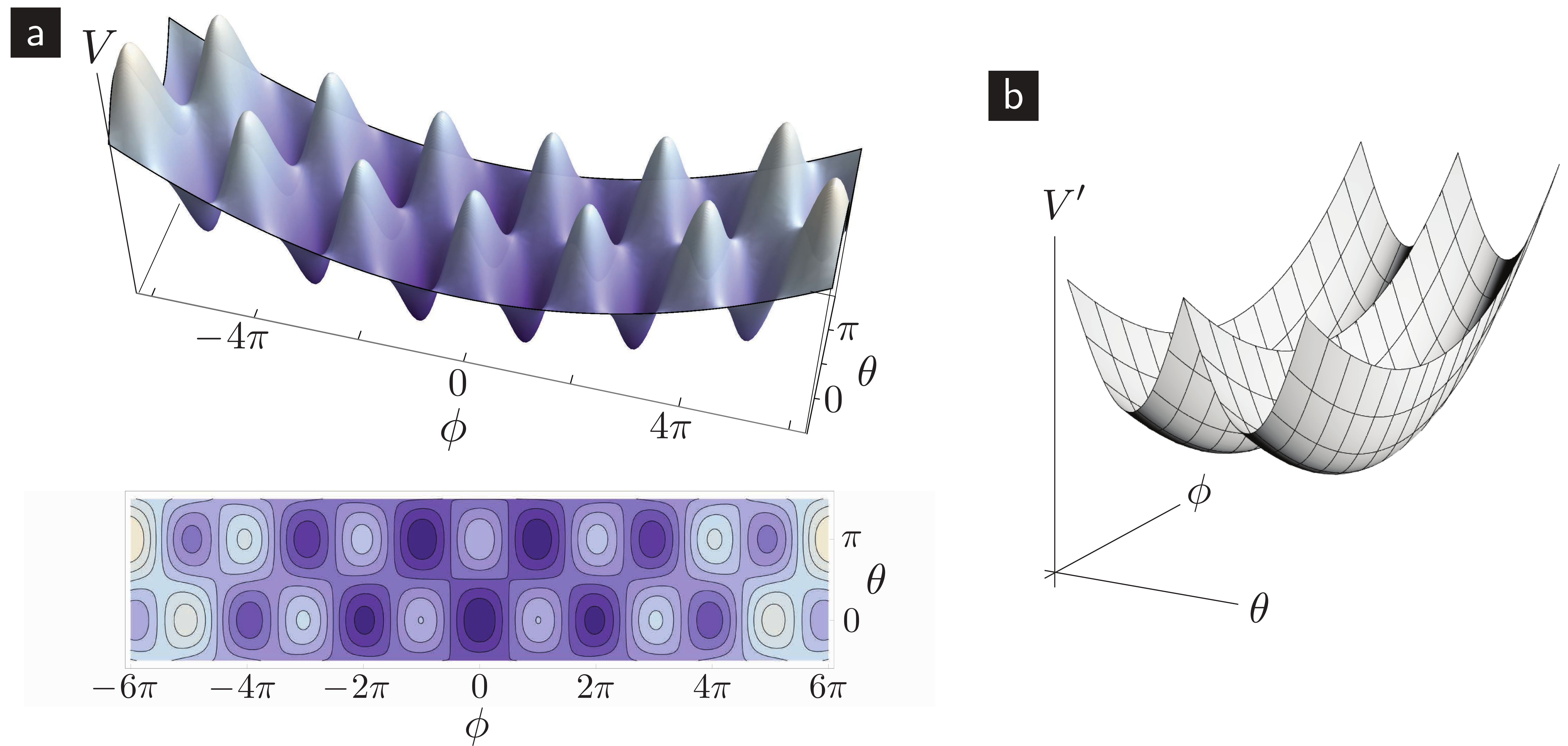}
	\caption{(Color online) (a) Potential energy $V(\phi,\theta)$, showing the two-fold fluxonium-like potential for cuts along the $\theta = 0$ and $\theta=\pi$ ridges. Localization wave function in these ridges occurs when the effective mass along the $\theta$ direction is sufficiently large. [Parameters: $\ej/E_L=165$.]
	(b)  Simplified model with separable potential energy. In this case, wave functions are products $\psi(\phi,\theta)=\psi_\text{ho}(\phi)\,\psi_\text{dw}(\theta)$ of harmonic-oscillator and double-well wave functions along $\phi$ and $\theta$ direction, respectively.
	\label{fig-potential}}
\end{figure*}

Carrying out the usual Legendre transform and canonical quantization, we finally arrive at 
\begin{align}\label{H}
H_\text{sym}=&-2\ecj\partial_\phi^2-2\ecs\partial_\theta^2\\\nonumber
&-2\ej\cos\theta\cos(\phi-\px/2)+E_L\phi^2 +2\ej.
\end{align}
as the Hamiltonian of the symmetric $0-\pi$ qubit.
The additional energy shift $2\ej$ included in $H$ is convenient in rendering the energy spectrum strictly positive. All charging energies in $H$ refer to the charge of a single electron so that $\ecj=e^2/2\cj$ and $\ecs=e^2/2\cs$. The potential energy $V(\phi,\theta)$ associated with the symmetric $0-\pi$ Hamiltonian is depicted in Fig.\ \ref{fig-potential}(a). The boundary conditions associated with $H_\text{sym}$ consist of square-integrability of the wave functions $\Psi(\phi,\theta)$  along the  real $\phi$ axis and $2\pi$ periodicity in the $\theta$ variable. 

\section{Spectrum of the symmetric $0-\pi$ device\label{sec:idealspectrum}}
\subsection{Qualitative discussion}
The effective potential for the $0-\pi$ circuit derived in Eq.\ \eqref{H}, $V(\phi,\theta)=-2\ej\cos\theta\cos(\phi-\px/2)+E_L\phi^2 +2\ej$, is shown in Fig.\ \ref{fig-potential}(a). For a qualitative understanding of low-lying eigenstates of the $0-\pi$ device, it is useful to consider a much simpler potential first, taking the form
\be
V'(\phi,\theta) = V_\text{dw}(\theta) + V_\text{ho}(\phi).
\ee
Here, $V_\text{dw}$ is a symmetric double-well potential and $V_\text{ho}$ a shallow harmonic oscillator potential as shown in Fig.\ \ref{fig-potential}(b). Due to the special form of this potential, the problem becomes separable and wave functions are products $\psi(\phi,\theta)=\psi_\text{ho}(\phi)\,\psi_\text{dw}(\theta)$ of harmonic-oscillator and double-well wave functions along the $\phi$ and $\theta$ coordinate, respectively. The two lowest-lying eigenstates correspond to Gaussian wave functions along $\phi$ and the symmetric and anti-symmetric superposition of states localized close to the the two double-well minima along the $\theta$ direction. Degeneracy of these two states is only weakly broken by tunneling  as long as the effective mass along the $\theta$ direction is heavy enough to suppress large fluctuations. 

\begin{figure*}
\centering
	\includegraphics[width=0.8\textwidth]{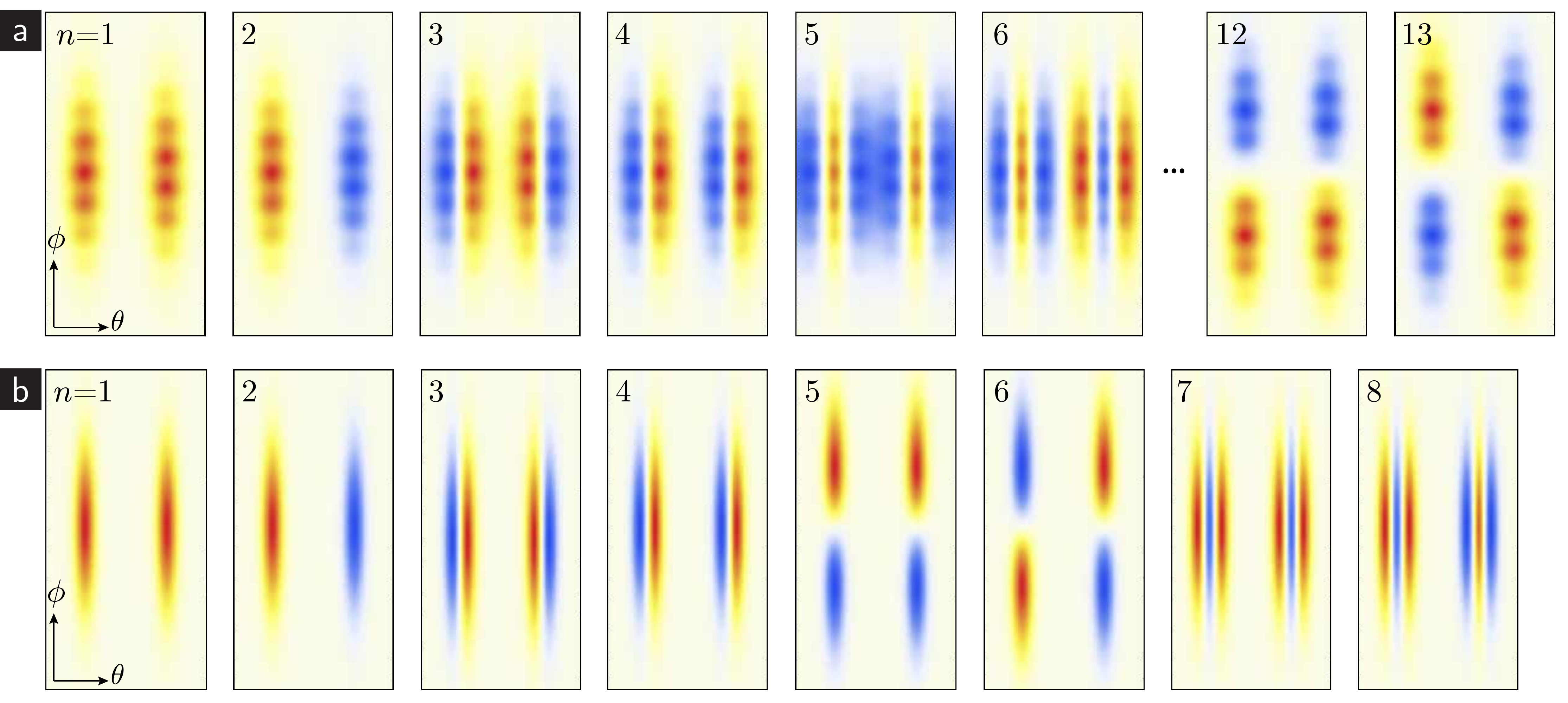}
	\caption{(Color online) Density plots of the wave function amplitudes for eigenstates in (a), the full potential $V$ of the $0-\pi$ qubit and (b), the separable potential $V'$. The numbers $n=1,2,\ldots$ enumerate the eigenstates starting from the ground state. Different colors (shades of gray) mark distinct signs of the wave function amplitudes. In the simpler case (b), localization along the two ridges $\theta=0$ and $\theta=\pi$ and pairing of states into doublets of symmetric and anti-symmetric states (in $\theta$ direction) are easily visible. Wave functions of the actual $0-\pi$ circuit in (a) show additional local extrema due to the cosine corrugation of the potential $V$. Overall comparison -- in particular, states $n=5$ and $6$ -- shows that delocalization in $\theta$ occurs more easily for the $0-\pi$ circuit. As a result, the development of nodes in $\phi$ direction (states 12 and 13) only takes place at higher energies.   [Parameter values: $\hbar\omega_p/E_L=10^4,\,\hbar\omega_p/\ecs=2.2\cdot10^3,\, \hbar\omega_p/\ej=7.9$.]}
        \label{fig:wvf}
\end{figure*}

As long as tunneling in the $\theta$ direction remains suppressed, excited states above these lowest two states will appear in doublets. Except for the small tunnel-induced splittings within doublets, level spacings in the spectrum will exhibit two separate energy scales: the harmonic-oscillator energy spacing from $V_\text{ho}$ and the spacing of states in each local-well of $V_\text{dw}$. When considering the form of the wave functions, the two energy spacings are associated with either an increase in the node number in $\phi$ direction or in $\theta$ direction. An example for the choice $V_\text{dw}(\theta)=-2\ej|\cos\theta|$ is shown in Fig.\ \ref{fig:wvf}(b), illustrating the localization along the two ridges $\theta=0$ and $\theta=\pi$ as well as the progressive increase in the number of nodes along the two directions.

We next consider the actual potential energy $V$ of the symmetric $0-\pi$ circuit, as shown in Fig.\ \ref{fig-potential}(a) for zero magnetic flux. Like the simplified  potential $V'$, the true potential $V$ has two ridges at $\theta=0$ and $\theta=\pi$  -- but additionally has oscillatory terms along the $\phi$ direction. While each of them resembles a fluxonium potential \cite{Koch2009,Manucharyan2009}, the minima in the $\theta=0$ ridge versus the  $\theta=\pi$ ridge are staggered with respect to one another (thus preventing separability). With the appropriate choice of circuit parameters, the interesting ground state degeneracy seen in the simplified toy model is also reflected in the physics of the actual circuit with the more complicated potential.

\subsection{Discussion of conditions for degeneracy}
A prototype for a single-particle nearly degenerate system is a  double well. In the mentioned toy model, it is clear that if the valleys are symmetric and tunneling is suppressed, then the states of the particle living in the left and right valley will be approximately decoupled and degenerate. 
For certain circuit parameters, the degeneracy in the $0-\pi$ circuit is very similar in nature. It will become apparent in Section \ref{sec:disorder} that the degeneracy of the $0-\pi$ circuit is especially robust against disorder.

When tunneling in the $\phi$ direction (i.e., in the direction along each ridge) is much larger than tunneling in the $\theta$ direction (from one ridge to the other), then the maxima in the $0-\pi$ potential can largely be ignored and one expects a similar degeneracy as in the toy model.
This difference in tunneling strengths can be achieved by  choosing significantly different effective masses along the $\phi$ and $\theta$ directions, namely $\ecj\gg\ecs$ (or, equivalently, $\cj\ll \cp$).
Localization along $\theta$ within each ridge is further strengthened by reducing the oscillator length for harmonic fluctuations along the $\theta$ direction, which is accomplished when $\ej\gg \ecs$.

The symmetry between the two ridges is broken because the minima of the two ridges are staggered with respect to the harmonic potential. Further, magnetic flux shifts both ridges with respect to the harmonic potential, leading to energy offsets. 
The sensitivity to both of these effects is reduced when wave functions are delocalized over multiple minima of the cosine potential within each ridge. This occurs when the parabolic envelope of the potential is sufficiently shallow, i.e., $E_L\ll\ej$, and the nominal oscillator length in the quadratic potential is large, i.e., $E_L\ll\ecj$.

Intuition for this insensitivity of states delocalized in $\phi$ is similar to the flux insensitivity of the fluxonium circuit \cite{Koch2009,Ferguson2013}. There, low-lying wave functions form metaplasmon states delocalized across multiple potential minima. These states are exponentially insensitive, with an exponential suppression factor of $\sim\nolinebreak\exp(-r\sqrt{\ecj/E_L})$ where $r>0$ is of order unity for low-lying levels \cite{Koch2007a,Koch2009}. We show that the same physics leads to  degenerate states insensitive to magnetic flux and  energy offsets in the $0-\pi$ circuit.

To summarize, we find that robust ground state degeneracy (up to exponentially small deviations) requires the following set of inequalities among device parameters to hold:
\begin{align}
E_L,\,\ecs \ll \ej, \, \ecj.
\label{degen params}
\end{align}

\subsection{Numerical Results for Wave Functions and Energy Levels}
Due to the cosine modulation along the $\phi$ direction and coupling between motion in $\phi$ and $\theta$ direction, the full potential $V$  of the $0-\pi$ circuit is not separable. We thus solve the corresponding Schr\"odinger equation  numerically to obtain energy levels and eigenstates. Specifically, we employ the finite-difference method in its simplest implementation (see Appendix \ref{app:finite}). With this method we can find the full solution in both the limit described in Eq.\ \eqref{degen params} but also in intermediate regimes where no clear hierarchy of energy scales exists.

Figure \ref{fig:wvf}(a) illustrates the resulting wave functions for the $0-\pi$ device deep in the degeneracy regime. Qualitative similarities with the wave functions of the simplified potential [Fig.\ \ref{fig:wvf}(b)] are evident. Important differences between the two cases include the additional structure of wave functions of the $0-\pi$ device brought on by the cosine corrugation of the potential, as well as an increased tendency of wave functions to spread in $\theta$ direction. The latter is easily understood from inspection of the potential $V$, showing that the two ridges are not separated by a large potential barrier along $\theta=\pi/2$. Nevertheless, the wave functions shown in Fig.\ \ref{fig:wvf} are qualitatively similar between the toy model and the actual potential.

\begin{figure*}
\centering
	\includegraphics[width=0.9\textwidth]{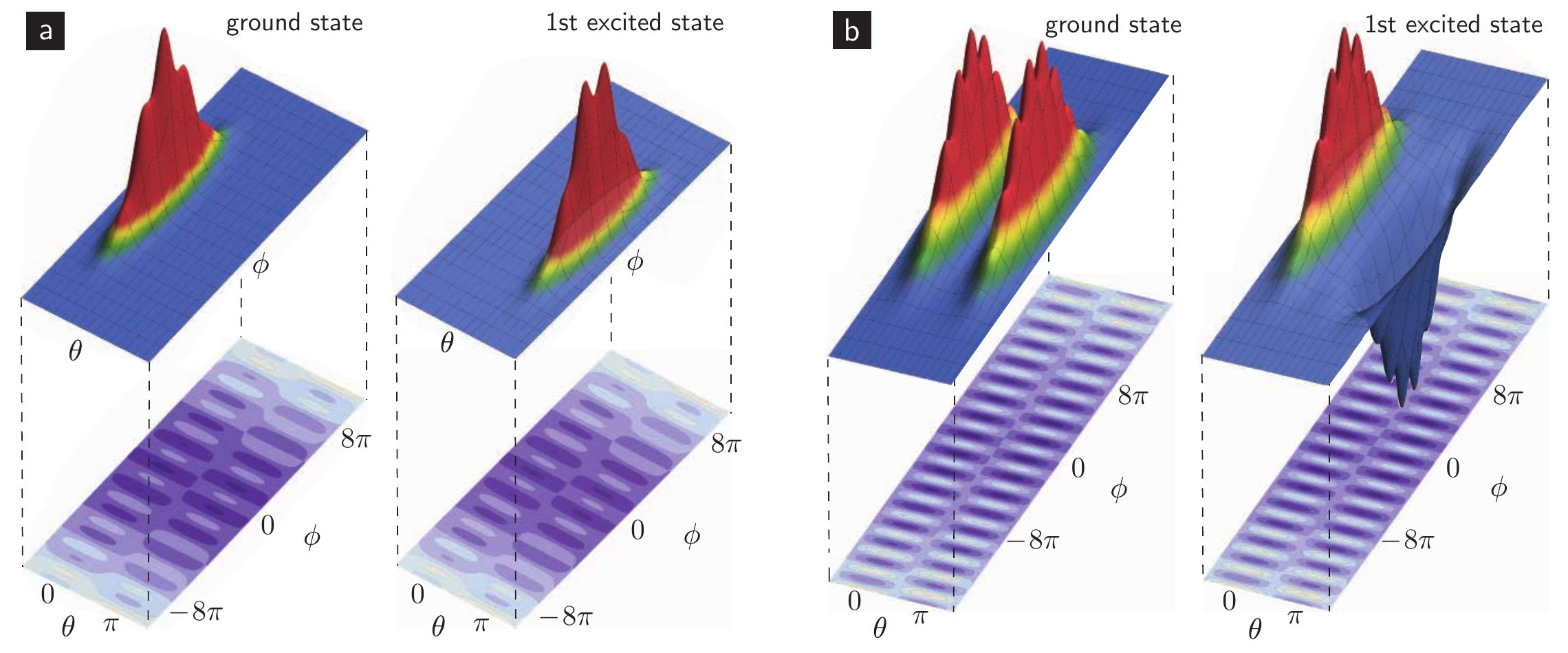}
	\caption{(Color online) Wave functions for the ground state and its (nearly) degenerate partner state for two choices of device parameters. While the value of the degeneracy is identical in the two cases ($D = 2.7$), it is limited by potential energy differences in the two ridges in (a), and by tunneling along the $\theta$ direction in (b). For tunneling-induced degeneracy breaking (larger $\ecs$ and smaller $E_L$), we observe symmetric and antisymmetric superpositions of the states localized in the individual ridges. For degeneracy breaking due to potential offsets between the two ridges (smaller $\ecs$ and larger $E_L$), we observe localization in the two separate ridges.	[Parameter values: (a) $\hbar\omega_p/E_L=9.9\cdot 10^2,\,\hbar\omega_p/\ecs=10^4,\, \hbar\omega_p/\ej=8.3$; (b) same as in Fig.\ \ref{fig:wvf}, i.e., $\hbar\omega_p/E_L=10^4,\,\hbar\omega_p/\ecs=2.2\cdot10^3,\, \hbar\omega_p/\ej=7.9$.]
	}
	\label{fig-wavefunction}
\end{figure*}

Which of these two types is formed generally depends on the parameters $E_L$, $\ecs$, and magnetic flux. As long as the magnetic flux is away from half-integer flux quanta, the staggering of local minima leads to an effective energy offset between the two ridges. Just as for an asymmetric double-well potential these energy offsets promote localization in the individual ridges, becoming more pronounced as $E_L$ is increased and leading to a ground state doublet of the type shown in Fig.\ \ref{fig-wavefunction}(a). 
Conversely, decreasing the effective mass along the $\theta$ direction, i.e., increasing $\ecs$ promotes tunneling and delocalization of the wave function, leading to eigenstates in the form of symmetric and anti-symmetric superpositions, as shown in Fig.\ \ref{fig-wavefunction}(b). In summary, by tuning the relative strength between $E_L$ and the tunneling (via $\ecs$), we can favor one type over the other.

\begin{figure}
\centering
	\includegraphics[width=0.85\columnwidth]{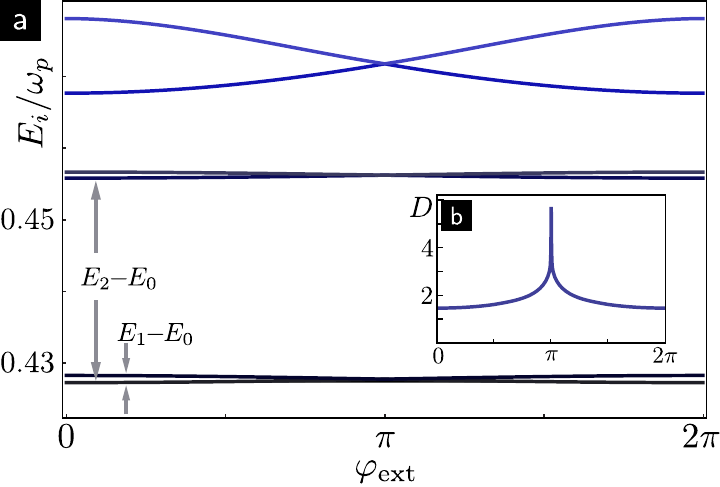}
	\caption{(Color online) (a) Energy spectrum as a function of external magnetic flux $\varphi_\text{ext}=\Phi_\text{ext}/\Phi_0$. For the selected parameters, the lifting of the degeneracy at zero flux is primarily induced by the potential asymmetry and strongly suppressed for $\varphi_\text{ext}=\pi$ where the potential becomes symmetric. 
	(b) Magnetic flux dependence of the logarithmic degeneracy parameter $D$. [Parameter values: $\hbar\omega_p/E_L=10^3$, $\hbar\omega_p/E_{\text{C}\Sigma}=10^3$, and $\hbar\omega_p/\ej=3.95$.]}
        \label{Spectrum}
\end{figure}

In principle, magnetic flux can also be used to generate superposition-type states: tuning $\Phi_\text{ext}$ to a half-integer flux quantum produces a potential which is symmetric with respect to the two ridges and, thus, does not exhibit an effective energy offset. 
Figure \ref{Spectrum} gives an example of the full flux dependence of low-lying energy levels. The doublet structure of the lowest four energy states is clearly visible, as is the suppression of the energy splitting at half-integer flux ($\px=\pi$). The metaplasmon-like character of the wave functions explains the relative  insensitivity of low-lying energy levels to the external magnetic flux.

We next assess the degree of degeneracy that can be achieved with realistic device parameters. To quantify the degeneracy we define the parameter $D$ by
\begin{align}
D=\log_{10}\frac{E_2-E_0}{E_1-E_0}
\end{align}
where $E_0,E_1$ and $E_2$ are the eigenenergies arranged in increasing order, starting with the ground state. $D$ thus specifies the ratio between the doublet energy splitting and the energy difference to the next higher doublet on a log scale, as illustrated in Fig.\ \ref{Spectrum}. In the absence of magnetic flux, wave functions in this example are of the type shown in Fig.\ \ref{fig-wavefunction}(a). The degeneracy $D$ is seen to reach a maximum at half-integer flux, which eliminates the energy offsets between the two ridges and switches to wavefunctions of the Fig.\ \ref{fig-wavefunction}(b) type. 
The following discussion will investigate the degeneracy at zero flux, away from the special flux value, to better highlight the interplay between the two regimes.

\begin{figure}
\centering
	\includegraphics[width=0.8\columnwidth]{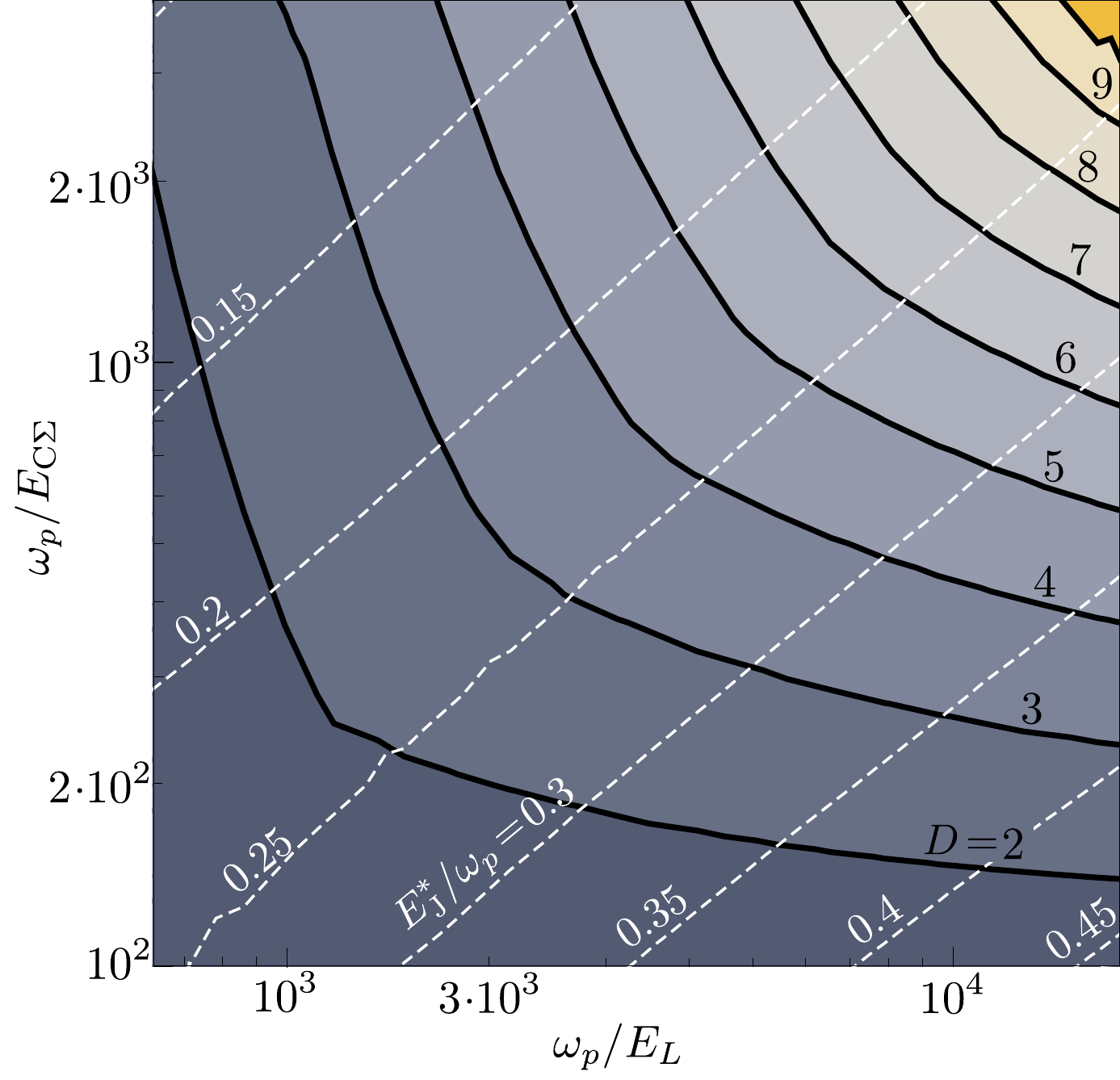}
	\caption{(Color online) Maximum value of the degeneracy parameter $D$ for given $E_L$ and $\ecs$ and optimal $\ej^*$. Contours of maximum $D$ are shown in black. Contours for the optimal values $\ej^*$ maximizing $D$ are shown as white dashed lines. Results underline that strong degeneracy requires challengingly small values of $E_L$ and $\ecs$.}
	\label{fig-dmax}
\end{figure}

An important question is the quantitative dependence of the degeneracy $D$ on the device parameters of the $0-\pi$ circuit. Indeed, the inequalities from Eq.\ \eqref{degen params} specify general requirements for finding near-degenerate pairs of low-lying states; however, Eq.\ \eqref{degen params} does not provide a concrete parameter range. To obtain this range, we systematically calculate the degeneracy $D$ for a large set of parameter choices as follows. We first note that variations in junction capacitance $\cj$ and Josephson energy $\ej$ are routinely achieved with Al-AlOx-Al junctions by changing the junction area while keeping the insulator thickness constant. Under these circumstances, the effective plasma frequency $\omega_p = \sqrt{8\ej\ecj}/\hbar$ remains fixed. We thus take  $\hbar\omega_p$ as our energy scale and treat $E_L$, $\ecs$, and $\ej$ as independent parameters; the junction capacitance takes the form $\ecj/\hbar\omega_p=\hbar\omega_p/8\ej$. We then form a logarithmic grid in the parameter plane spanned by $E_L$ and $\ecs$. For each grid point, we calculate the degeneracy $D$ and finally vary $\ej$ to find the maximum degeneracy value $D_\text{max}$ (for given $E_L$ and $\ecs$). Our key results are depicted in the log-log plot shown in Fig.\ \ref{fig-dmax}.

The constant-$D$ contours in Fig.\ \ref{fig-dmax} illustrate that there are indeed two qualitative regimes for reaching high degeneracy values, which is fully consistent with the two types of doublet states shown in Fig.\ \ref{fig-wavefunction}. Whenever $E_L$ is sufficiently small, the degeneracy $D$ is mainly limited by the splitting induced by tunneling along the $\theta$ direction. Accordingly, $D$ can be increased by further suppressing tunneling, as is achieved by decreasing the value of $\ecs$. In this regime, wave functions are symmetric and anti-symmetric superpositions of wave functions localized in the $\theta=0$ and $\theta=\pi$ ridges, as shown in Fig.\ \ref{fig-wavefunction}(b). Vice versa, when $\ecs$ is sufficiently small, $D$ is predominantly governed by the asymmetry between the two potential ridges. This asymmetry can be lowered by decreasing the superinductance energy $E_L$. Wave functions in this regime are localized in one ridge or the other, see Fig.\ \ref{fig-wavefunction}(a). 

In Fig.\ \ref{fig-dmax}, we also show the contours for $\ej^*$, defined as the value of $\ej$ that maximizes the degeneracy for given $E_L$ and $\ecs$. These contours are, approximately, straight and parallel lines with a unit slope, implying that the optimum values obey the parametric dependence $\ej^*=f(\ecs/E_L)$. Contours of $\ej^*$ are nearly equidistant, implying that $\ej^*$ is nearly a plane surface in log-log space over the investigated parameter range. Numerically, we find the approximate relation
\be
\frac{\ej^*}{\hbar\omega_p}\approx 0.17 - 0.11\cdot\log_{10} (\ecs/E_L).
\ee 

To illustrate the practical challenge in reaching the parameter regime of near degeneracy, we consider relatively moderate values of $\omega_p/\ecs=\omega_p/E_L=10^3$ for the relevant charging and superinductance energies, which achieves a degeneracy value of $D\approx2$ according to Fig.\ \ref{fig-dmax}. 
For a device with a plasma oscillation frequency of the order of $\omega_p/2\pi=40\,\text{GHz}$ (typical of Al-AlOx junctions), this implies a superinductance of roughly $4\,\mu\text{H}$ and a capacitance $C$ of about $1\,\text{pF}$. Experimentally realized superinductances are approaching the threshold value necessary for reaching the robustly degenerate regime \cite{Bell2012,Masluk2012}.

\section{Effects of Disorder\label{sec:disorder}}
Unavoidable device imperfections will generally lead to some amount of disorder in the parameters of the $0-\pi$ circuit. Specifically, the parameters of each pair of junctions, capacitors and superinductors in the circuit will not be precisely identical. 
We thus consider the effect of such disorder on the spectrum  of the $0-\pi$ circuit and on the degeneracy $D$, in particular. 

When including parameter disorder, the kinetic and potential energies [previously Eqs.\ \eqref{kinetic}--\eqref{potential}]  take the more general form
\begin{align}
T=\bcj\dot\phi^2 + (\bcp+\bcj)\dot\theta^2+\bcp\dot\chi^2
+2\bdcj  \dot \phi \,\dot\theta
+2\bdc \dot\theta\, \dot\chi
\end{align}
and
\begin{align}\nonumber
U=&-2\ej \cos(\theta)\cos(\phi-\tfrac{1}{2}\px) +2\dej \sin(\theta)\sin(\phi-\tfrac{1}{2}\px)\\
&+E_L \phi^2+E_L\chi^2+2\del \phi \chi.
\label{full potential}
\end{align}
Here,  $C=(C_1+C_2)/2$ now denotes the arithmetic mean of the two capacitors and $\delta C=(C_1-C_2)/2$ the deviation from the mean. We employ analogous definitions for disorder in the various other circuit parameters. 

A Legendre transform and subsequent series expansion in the capacitive disorder then leads to the Hamiltonian
\begin{align}
\nonumber
H&\simeq H_\text{sym}
+4\ecs(\dcj/\cj)\partial_\phi\partial_\theta +2\,\dej \sin\theta\sin(\phi-\px/2)\\
&\quad-2\ec\partial_\chi^2+E_L \chi^2+4\ecs(\dc/\cp)\partial_\theta\partial_\chi+2\,\del \phi\, \chi, 
\label{full potential}
\end{align}
where we have dropped contributions $\sim\mathcal{O}(\dc^2,\dcj^2,\dc\,\dcj)$.
The terms in the first line of Eq.\ \eqref{full potential} comprise the previous model of the symmetric $0-\pi$ device [Eq.\ \eqref{H}] plus small corrections due to disorder in the parameters $\cj$ and $\ej$ describing the two Josephson junctions. The second line contains the harmonic terms for the $\chi$ degree of freedom, as well as two  terms from disorder in $E_L$ and $\cp$ which couple between the $\chi$ degree of freedom and the fundamental $0-\pi$ circuit variables  ($\phi$, $\theta$). In the following, we discuss the effects of these different types of disorder.

\begin{figure*}
\centering
	\includegraphics[width=.95\textwidth]{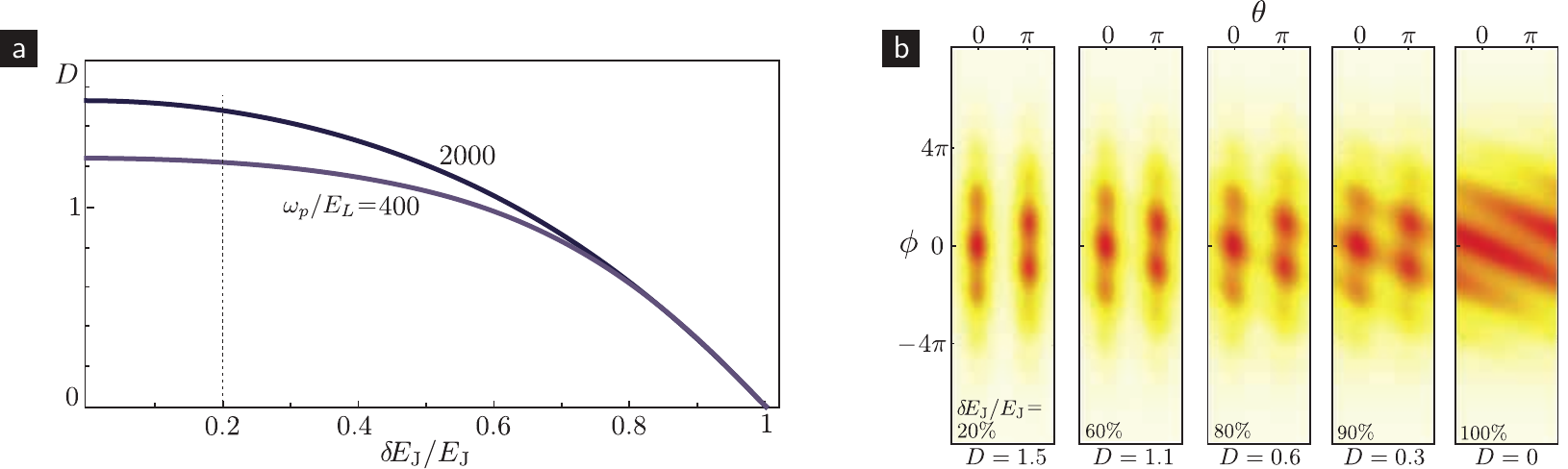}
	\caption{(Color online) Effect of disorder in the Josephson energies. (a) Dependence of the degeneracy parameter $D$ on relative disorder in the Josephson energy,  $\dej/\ej$. The plot shows a comparison of two different parameter sets, both with $\hbar\omega_p/E_{\text{C}\Sigma}=10^3$ and $\hbar\omega_p/\ej=7.9$ and zero magnetic flux. The degeneracy is seen to be fairly robust with respect to $\ej$ disorder. The vertical line at $20\%$ disorder marks the worst-case disorder seen in experiments. (b) Density plots of the ground state wave function showing the expected deformation as $\ej$ disorder is increased.}
        \label{DisorderEJ}
\end{figure*}

\subsection{Disorder in junction parameters $\ej$ \emph{and} $\cj$}
Disorder in the Josephson junction parameters ($\dej$ and $\dcj$) is straightforward to incorporate as it does not introduce coupling between the fundamental $0-\pi$ circuit variables and the additional harmonic degree of freedom captured by the $\chi$ variable. The effects of junction disorder can thus be treated by the same numerical diagonalization scheme as before.

As seen from Eq.\ \eqref{full potential}, disorder in the junction capacitance $\cj$ only leads to a slight change in the effective mass tensor. Corrections due to this are expected to be small since $\ecs\,\dcj/\cj < \ecs \ll \ecj$. The critical condition for maintaining robust degeneracy in the presence of $\cj$ disorder, is that the tunneling along $\phi$ must remain strong and tunneling along $\theta$ must remain weak. 
As long as the $\dcj$ (and $\cj$) remains small compared to $C$, this tunneling condition will still be satisfied. Indeed, results from numerics show that the effect of this disorder  is negligible for values up to $\dcj/\cj=100\%$. This should be compared to the conservative estimate of experimental disorder in $\cj$ of up to $10\%$, mainly caused by edge imperfections in the double-angle evaporation used for the fabrication of Al-AlOx Josephson junction.

Disorder in the Josephson energies leads to a distortion of the potential energy $V(\phi,\theta)$. According to Eq.\ \eqref{full potential}, this distortion is directly proportional to $\dej$ and can hence produce noticeable changes in wave functions, eigenenergies and the degeneracy measure $D$.  Representative numerical results are shown in Figure \ref{DisorderEJ}. The degeneracy  $D$  is fairly robust for realistic amounts of $\ej$ disorder [Fig.\ \ref{DisorderEJ}(a)]. Experimentally, Josephson energies are known to vary from device to device by up to $20\%$; disorder among junctions within the same device are expected to be significantly smaller than this. The reason for the rapid drop of $D$ at very strong disorder is illustrated in Fig.\ \ref{DisorderEJ}(b), showing the dramatic change of wave functions as the potential energy is more and more deformed. Strong $\ej$ disorder eliminates the two potential ridges along $\theta=0$ and $\theta=\pi$ and wave functions spread over the full range along the $\theta$ direction. Consequently, $\ej$ disorder ultimately destroys the degeneracies of low-lying states -- however, only for disorder strengths that vastly exceed the amount of disorder expected in experiments.

\subsection{Disorder in $\cp$ and $E_L$}
\begin{figure*}
\centering
	\includegraphics[width=0.9\textwidth]{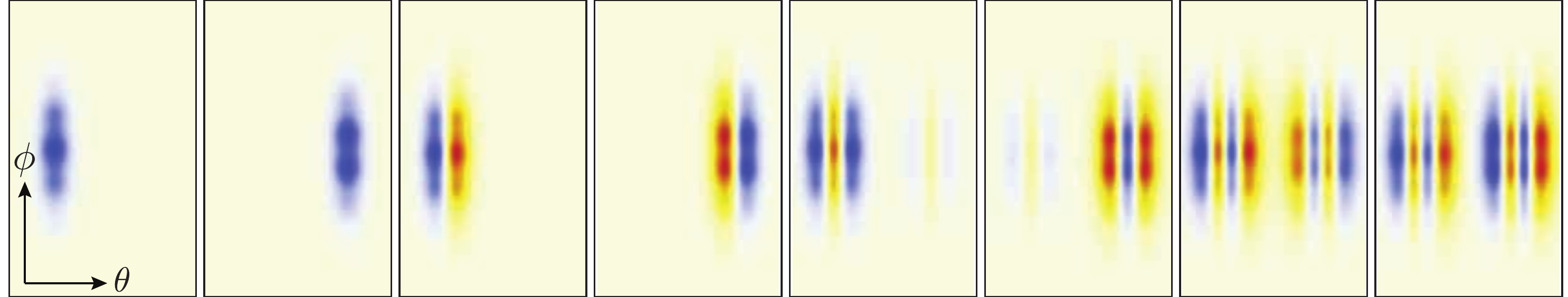}
	\caption{(Color online) Density plot of low-lying wave functions with disjoint support. [Same parameters as in Fig.\ \ref{fig-wavefunction}(b).]}
        \label{fig-disjoint}
\end{figure*}

Both disorder in the capacitance $\cp$ as well as in the superinductance energy $E_L$ introduce coupling between the $0-\pi$ device variables ($\phi,\theta$) and the harmonic variable $\chi$. This is similar to the typical situation of circuit QED where a qubit is coupled to a harmonic oscillator, and can thus be treated by the same methods \cite{Blais2004}. In the eigenbasis $\{\ket{l}\}_{l=0,1,\ldots}$ of the symmetric $0-\pi$ circuit [Eq.\ \eqref{H}], the full Hamiltonian can be rewritten as
\be\label{H2}
H=\sum_l E_l^\text{sym}\ket{l}\!\bra{l} + \hbar\Omega_\chi a^\dag a
+\sum_{l, l'}\left( g_{ll'}\ket{l}\!\bra{l'}a +\text{h.c.}\right)
\ee
where $g_{ll'}=g^\phi_{ll'}+i g^\theta_{ll'}$ are coupling strengths defined by 
\begin{align}\label{g}
g_{ll'}^\theta &= \ecs (\dc/\cp) \,(32E_L/\ec)^{1/4}\,\boket{l}{i\,\partial_\theta}{l'},\\
g_{ll'}^\phi &= \del\,(8\ec/E_L)^{1/4} \boket{l}{\phi}{l'},
\end{align}
and $a$ ($a^\dag$) is the annihilation (creation) operator for excitations of the $\chi$ oscillator. Due to disjoint support of wave functions as well as parity, we expect certain instances of the occurring matrix elements to be strongly suppressed, see Fig.\ \ref{fig-disjoint}.

Following the general approach from Ref.\ \onlinecite{Zhu2013}, we obtain the dispersive Hamiltonian
\be
H'=\sum_{l=0}^\infty (E_l^\text{sym}+\kappa_l)\ket{l}\!\bra{l} + \hbar\Omega_\chi a^\dag a +\sum_{l} \chi_l \ket{l}\!\bra{l}a^\dag a
\ee
where
\be
\chi_l=\sum_{l'}\abs{g_{ll'}}^2\left(\frac{1}{\Delta_{ll'}}-\frac{1}{\Delta_{l'l}}\right), \qquad
\kappa_l=\sum_{l'}\frac{\abs{g_{ll'}}^2}{\Delta_{ll'}}
\ee
are the ac Stark shift and the Lamb shift, respectively. The detuning is defined as 
$\Delta_{ll'}=E_l^\text{sym}-E_{l'}^\text{sym}-\hbar\Omega_\chi$. We note that if there are resonances between the $0-\pi$ circuit and the harmonic oscillator, this perturbative treatment may break down. For small disorder, we expect that the Lamb shifts $\kappa_l$ will be small compared to the splitting between each doublet.

\section{Concluding remarks\label{sec:conclusions}}
We have developed a full circuit analysis of the $0-\pi$ superconducting circuit, which is valid both in the highly degenerate regime as well as in intermediate parameter regimes where both the ground state and low-lying excited states are important.  We find that in the case of symmetric parameter values and no disorder, the system can be decomposed into an uncoupled harmonic degree of freedom, and a subsystem subject to a two-dimensional effective potential.  In a certain regime, the spectrum of this subsystem consists of degenerate doublets whose ground state splitting is exponentially small compared with the spacing between the lowest two doublets ($\approx \omega_p \sqrt{\ecs / \ecj}$).  If such degenerate states could be utilized as quantum bits, they would be protected from both dephasing and relaxation and not require fine-tuning of parameters.  However, realizing universal operations on such states is not a trivial task, and is still a subject of active inquiry.  Reaching the degenerate regime of $D>2$ (a 100-fold suppression) requires realizing inductances only slightly larger than the current state-of-the-art, which seems possible with continued advances in design and microfabrication techniques.  
 
A careful study of the effects of disorder has been presented.  Deep within the degenerate regime, even large disorder  in  the circuit parameters will not strongly affect the degeneracy.  In intermediate regimes, it is clear that disorder becomes important, especially disorder which introduces coupling to the $\chi$ harmonic mode. We have shown that this coupling can be treated using the formalism of circuit QED.  Future work will consider the potential to exploit this additional quantum degree of freedom for readout and manipulation, as well as to understand the effect of thermal fluctuations.
 
The $0-\pi$ circuit and the analysis method we have employed point to a new strategy of engineering the potential and kinetic energies of circuits with larger numbers of degrees of freedom: the realization of protected manifolds, suitable for quantum information processing, through the design of potential landscapes with specific properties.  Even in the absence of strict degeneracy, the presence of doublet $\lambda$ systems in the energy spectrum represents a promising route to realizing ultra-coherent qubits.  Finally, it may be possible to employ more complex circuit topologies in the future design of potential energy landscapes.

\begin{acknowledgments}
We thank Andrew Houck, Jay Lawrence, Andy C.\ Y.\ Li, David McKay, Thomas Yu, and Guanyu Zhu for stimulating discussions.  Our research was supported by the NSF under grants PHY-1055993 (JD, BF, DGF, JK), DMR-0805277 (DGF), DMR-1151839 (DIS),  by the LPS/NSA under ARO contract W911NF-12-1-0608 (DIS), by the Sloan Foundation (DIS) and the Packard Foundation (DIS).
\end{acknowledgments}

\appendix
\section{Numerical diagonalization of the $0-\pi$ Hamiltonian\label{app:finite}}
For numerical diagonalization of the $0-\pi$ circuit Hamiltonian, we employ the finite-difference method in its simplest possible form. We truncate $\phi$ to a finite interval $[-\phi_M,\phi_M]$ and discretize $\phi$ and $\theta$ according to $\phi_m = m\Delta_\phi$ ($m=0,\pm1,\ldots,\pm M)$ and $\theta_n=n\Delta_\theta$ ($n=1,\ldots,N$) so that  
$\theta_N=N\Delta_\theta=2\pi$. The corresponding orthonormal set $\{\ket{nm}\}$ of discretized position states is defined in the usual way by $\psi_{nm}(\phi,\theta)=\bket{\phi,\theta}{nm}=(\Delta_\phi\Delta_\theta)^{-1/2}$ whenever $(\phi,\theta)$ lies inside the rectangle centered at $(\phi_n,\theta_m)$ with width and length set by the grid constants $\Delta_\phi$, $\Delta_\theta$. Everywhere else, the wave function vanishes. 
For sufficiently fine grid, the matrix elements of potential energy and kinetic energy are approximated by using
$
\bra{m'n'}V(\phi,\theta)\ket{mn}\approx \delta_{mm'}\delta_{nn'}V(\phi_m,\theta_n)
$
and
\begin{align*}
&\bra{m'n'}\partial_\phi^2 \ket{mn} \approx \Delta_\phi^{-2}(\delta_{m',m+1}\delta_{n'n}+\delta_{m',m-1}\delta_{n'n}-2\delta_{m'm}\delta_{n'n}),\nonumber\\
&\bra{m'n'}\partial_\theta^2 \ket{mn} \approx \Delta_\theta^{-2}(\delta_{m'm}\delta_{n',n+1}+\delta_{m'm}\delta_{n',n-1}-2\delta_{m'm}\delta_{n'n}).
\end{align*}
With this, the stationary Schr\"odinger equation reduces to a sparse eigenvalue problem which we solve numerically, while carefully checking for discretization errors and convergence.

\end{document}